**Effects of phase-gradient on the nonadiabatic dynamics and photon-phonon conversion in one-dimensional array of optomechanical cavities**

**Divya Mishra and Parvendra Kumar**
*Optics and Photonics Centre, Indian Institute of Technology Delhi, Hauz Khas, New Delhi, 110016*
*parvendra@opc.iitd.ac.in*

Manipulation of photonic and phononic coupling in the coupled cavities plays a crucial role in the development of nonreciprocal devices and photon-phonon conversion. Here, we theoretically investigate how the phase gradient of a driving laser affects the coupling between photonic and phononic modes. This, in turn, affects and offers the controllability of the nonadiabatic dynamics of the population of hybrid eigenmodes following a sudden change in the optomechanical coupling. We show that the controllable nonadiabatic dynamics can be attributed to the phase-induced alteration of the bandgaps in the hybrid eigenmodes. We further demonstrate that the phase-assisted control of the relative weight of photonic and phononic modes in the hybrid eigenmodes also contributes to the nonadiabatic dynamics. Finally, we investigate the effects of phase on the coherent conversion of photon-phonon in a finite-size array.

## I. Introduction

Controlling the optomechanical coupling in cavity optomechanics has been key to a number of important theoretical and experimental developments, such as the Su-Schrieffer-Heeger (SSH) model, topological optomechanics, nonreciprocal devices, and quantum transduction [1–9]. The optomechanical coupling between a single photon and a single phonon, known as vacuum optomechanical coupling, can be controlled and enhanced by driving an optomechanical system with a continuous wave laser via radiation pressure force [10]. It is worth mentioning that an optomechanical system can be built on a number of different platforms, such as a microtoroid, a microdisk resonator, a suspended membrane, a micromechanical membrane in a superconducting circuit, or a photonic crystal nano beam cavity [ 11–15].

A one-dimensional array of optomechanical cavities (1DOMA), in particular, has been studied substantially for investigating the role of controllable optomechanical coupling in nonreciprocal transport, photon-phonon conversion, nonadiabatic dynamics and topological photonics with gauge fields [16-25]. The realization of most of these phenomena relies on the capability to manipulate and control the bandgaps [17-19]. In fact, A. Seif et al. recently showed that the phase-dependent optomechanical coupling causes an asymmetry in the bandgaps of eigenmodes, which enables the nonreciprocal flow of phonons [17]. Also, J. Cao et al. show that photon-phonon conversion can be controlled using edge states that are protected by topology in an optomechanical array [8]. The nonadiabatic dynamics produced by a sudden change of Hamiltonian's parameter elucidate the fundamentals of thermalization in a many-body system, quantum phase transition, and entanglement growth [26]. Besides the physical systems like cold atomic and 2D Fermi gases in tunable optical lattices, the nonadiabatic dynamics and phase transitions have also been investigated by changing the optomechanical coupling from positive to negative value in a 1DOMA [18, 27-29].

In this work, we present a detailed investigation on how the phase gradient of a continuous wave laser drive affects the nonadiabatic dynamics and the generation of new excitations. It is shown that following a sudden change of optomechanical coupling from positive to negative value, the temporal dynamics and net excitations in eigenmodes can be controlled by manipulating the phase of the laser drive. We clarify and show that these effects can be attributed to phase-induced alteration of the bandgaps and switching between the constituent photonic and phononic modes in the hybrid eigenmodes. Moreover, the effects of phase on the coherent photon-phonon are also investigated.

## II. Theoretical Model

The schematic of a one-dimensional array of optomechanical cavities is illustrated in Fig. 1. Here, each site represents a single optomechanical cavity which supports localized optical (blue circle) and mechanical (red circle) modes. The optomechanical coupling i.e. coupling between optical and mechanical modes is depicted by $G$. The hopping strengths between the two nearest optical and mechanical modes are represented by $J$ and $K$, respectively. Additionally, optical mode at each site is driven by a phase modulated laser with site dependent phase, $e^{in\theta}$. This position dependent phase breaks time-reversal symmetry [17] and introduces a position dependent effective coupling between optical and mechanical modes as shown in Fig. 1(b). It is important to note that each optomechanical site in the current model can be experimentally implemented using a photonic crystal nanobeam cavity on a silicon-on-insulator microchip [10]. The coupling of photonic and phononic modes between two sites can be accomplished by physically connecting them via a silicon beam segment, which acts as a waveguide for both optical and phononic modes [30]. Additionally, the hopping rates of photonic and phononic modes can be regulated by modifying the shape and number of holes in the nanobeam cavity [31, 32].

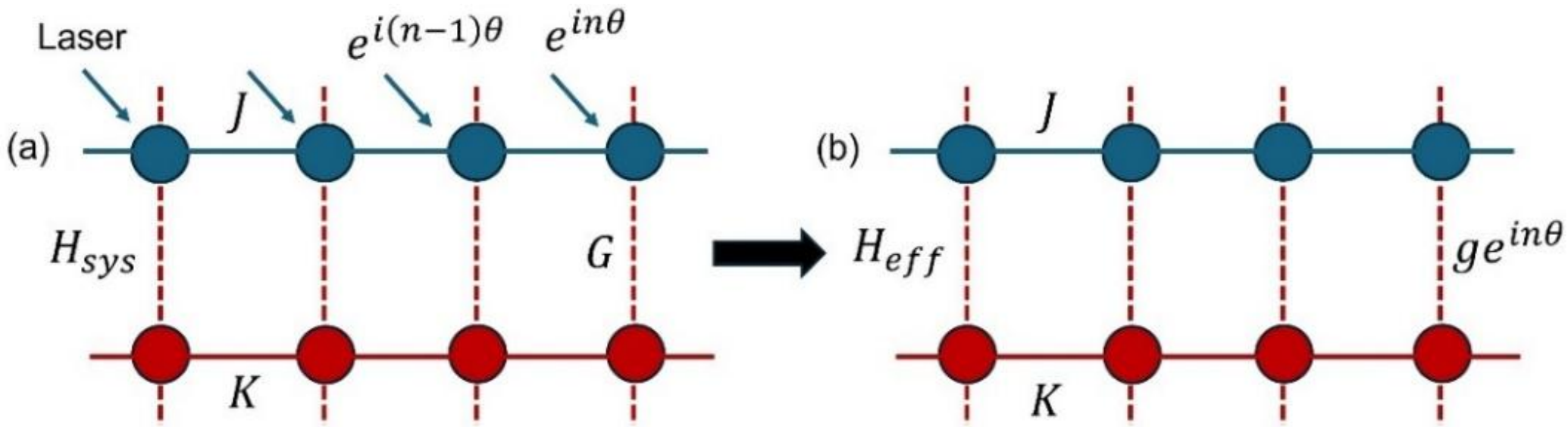

Fig. 1 (color online) Schematic of the optomechanical array. (a) Each site consists of a coupled optical mode (blue circle) and a phonon mode (red circle) with coupling strength $G$. The hopping strengths between nearest neighbour optical and phonon modes are represented by $J$ and $K$, respectively. The laser driving of optical modes with site dependent phase $e^{in\theta}$ transforms the bare Hamiltonian $H_{sys}$ into an effective Hamiltonian $H_{eff}$ with an enhanced and position dependent optomechanical coupling $ge^{in\theta}$ as shown in (b).

The site-dependent phase, as shown in Fig. 1(a), can be implemented using a multimode interference beam splitter and waveguides evanescently coupled to the nanobeam cavities, as described in Refs [17, 30]. Specifically, a multimode interference beam splitter can distribute the input power from a single laser uniformly among many output ports connected to waveguides of varying lengths, facilitating the generation of distinct phases according to the pathways. The Hamiltonian of OMA shown in Fig. 1(a) reads as [10, 17]:

$$H_{sys} = H_s + H_t, \tag{1}$$

$$H_s = \sum_n \omega_c a_n^\dagger a_n + \omega_m b_n^\dagger b_n - G a_n^\dagger a_n \left(b_n^\dagger + b_n\right) + \Omega_d cos(\omega_d t + n\theta)\left(a_n^\dagger + a_n\right), \tag{2a}$$

$$H_t = -J \sum_n \left(a_n^\dagger a_{n+1} + a_{n+1}^\dagger a_n\right) - K \sum_n \left(b_n^\dagger b_{n+1} + b_{n+1}^\dagger b_n\right) \tag{2b}$$

Here, $H_s$ and $H_t$ represent the Hamiltonians of each sites and tunnelling between them, respectively, $G$ represents a vacuum coupling rate. The bosonic operator $a_n$ ($b_n$) destroy a photonic (phononic) excitation with energy $\omega_c$ ($\omega_m$), whereas $\Omega_d$, $\omega_d$, and $\theta$ represent the Rabi frequency, frequency, and phase of the driving laser, respectively. We transform $H_{sys}$ in a frame rotating with angular frequency $\omega_d$. The linearized Hamiltonian in the rotating wave approximation reads as [10]:

$$H_{eff} = -\frac{\Delta}{2} \sum_n a_n^\dagger a_n + \frac{\omega_m}{2} \sum_n b_n^\dagger b_n - J \sum_n a_n^\dagger a_{n+1} - K \sum_n b_n^\dagger b_{n+1} - g \sum_n e^{-in\theta} a_n^\dagger b_n + h.c., \tag{3}$$

where $g = G\alpha$ is the optomechanical coupling strength enhanced by a factor of $\alpha = \sqrt{n_o}$, where $n_o$ is the average number of photons in the optical cavity mode, and $\Delta = (\omega_d - \omega_c)$. This reads as $\alpha = \Omega_d / \left[\left(\Delta_0 + \frac{i\kappa}{2}\right) + 2J cos\theta\right]$, $\Delta_0 = \Delta + G(\beta + \beta^*)$ and $|\beta|^2$ represents the average phonon number in the mechanical mode, where $\beta = G|\alpha|^2 / \left[\left(\omega_m - \frac{i\Gamma}{2}\right) - 2K\right]$, $\kappa$ and $\Gamma$ represent the decay rates of optical and mechanical modes, respectively. Note that the optomechanical coupling strength $g$ can be tuned by changing the intensity of the driving laser.

## III. Results and Discussions

To diagonalize the Hamiltonian in Eq. 3, we define the Fourier transform of $a_n$ and $b_n$ in terms of pseudomomentum operators $a_k$ and $b_k$ as $\begin{pmatrix} a_n \\ b_n \end{pmatrix} = \sum_k e^{-inkd} \begin{pmatrix} a_k e^{-in\theta} \\ b_k \end{pmatrix}$. The Hamiltonian in Eq. 3 in Fourier basis reads as

$$H_k = \begin{pmatrix} -\Delta - 2J\, cos(kd + \theta) & -g \\ -g & \omega_m - 2K cos(kd) \end{pmatrix}, \tag{4}$$

here, $k$ and $d$ represent the wavenumber and lattice constant, respectively. It is clear from Eq. 4 that the phase gradient shifts the momentum and facilitates the coupling of phonons and photons with different momenta. Furthermore, the change of basis is a valid approximation for sufficiently large number of lattices [17]. The eigenvalues of $H_k$ read as $\omega_k(\pm) = (\Omega - \xi)/2 \pm \sqrt{g^2 + \delta^2}$, where, $\Omega = \omega_m - 2K cos(kd)$, $\xi = \Delta + 2J cos(kd + \theta)$, and $\delta = (\Omega + \xi)/2$, whereas $A_k = \alpha_A a_k + \beta_A b_k$ and $B_k = \alpha_B a_k + \beta_B b_k$ are the corresponding eigenmodes. Clearly, both eigenmodes consist superposed optical and mechanical modes and therefore they are termed as hybrid modes. The hybridization of optical and mechanical modes arises from optomechanical interaction. The coefficients $|\alpha_A|^2 = g^2 / \left(g^2 + \left(\delta + \sqrt{g^2 + \delta^2}\right)^2\right)$ and $|\beta_A|^2 = \left(\delta + \sqrt{g^2 + \delta^2}\right)^2 / \left(g^2 + \left(\delta + \sqrt{g^2 + \delta^2}\right)^2\right)$ are the relative weights of photons and phonons in eigenmode $A_k$, while $|\alpha_B|^2 = g^2 / \left(g^2 + \left(\delta - \sqrt{g^2 + \delta^2}\right)^2\right)$ and $|\beta_B|^2 =$

$\left(\delta-\sqrt{g^2+\delta^2}\right)^2/\left(g^2+\left(\delta-\sqrt{g^2+\delta^2}\right)^2\right)$ are the relative weights of photons and phonons in eigenmode $B_k$. The transformation matrix relating two eigenmodes with the constituent modes is given as $\begin{pmatrix} A_k \\ B_k \end{pmatrix} = R_k \begin{pmatrix} a_k \\ b_k \end{pmatrix}$, where $R_k = \begin{pmatrix} \alpha_A & \beta_A \\ \alpha_B & \beta_B \end{pmatrix}$. Throughout this paper, we work in resolved sideband, $\kappa < \omega_m$ and strong coupling, $g > \kappa$ regimes. The simulation parameters, $\omega_m = 4.3\ GHz$, $\kappa = 43\ MHz$, $\Gamma = 4.3\ MHz$, are achievable with the current experiments, here $\kappa$ and $\Gamma$ represent the decay of photonic and phononic modes in the optomechanical cavity, respectively [17, 30]. We choose two distinct regimes in which the optomechanical coupling is lower or higher compared to the hopping strengths, i.e., $g < J, K$ and $g > J, K$, for investigating the nonadiabatic dynamics [30]. It is worth noting that we include the dissipation only for calculating the initial state but neglected in the nonadiabatic dynamics of eigenmodes, which is a valid approximation for $\kappa t_q \ll 1$, here $t_q$ is the quench time. This condition is perfectly satisfied for all considered parameters throughout this paper.

### A. Nonadiabatic dynamics of the eigenmodes

To study the nonadiabatic dynamics of eigenmodes, we model the time dependence in optomechanical coupling strength as $g(t) = g\left(1 - 2t/t_q\right)$ by following Ref. [18]. Here, $t_q$ is the quench time and specifies how quickly the coupling strength is changed in the Hamiltonian. The system dynamics, beginning from $t = 0$ to $t = t_q$, switches the coupling from $+g$ to $-g$. A large (small) value of $t_q$ describes a slow (fast) change of coupling and adiabatic (nonadiabatic) evolution. The boundary between the adiabatic and nonadiabatic evolution can be described by comparing the rate of quench with the bandgap energy for zero optomechanical coupling, $t_q \approx 2g/\Delta_{g=0}^2$, here $\Delta_{g=0} = J cos(kd + \theta) - K cos(kd)$ is the bandgap energy [18]. For $t_q \ll 2g/\Delta_{g=0}^2$, the evolution would be nonadiabatic and generate new excitations due to the population transition between the eigenmodes; on the contrary, the evolution would be simply adiabatic, and no extra excitations are generated for $t_q \gg 2g/\Delta_{g=0}^2$. Therefore, here, we investigate the role of phase in nonadiabatic dynamics of eigenmodes by setting the $t_q \ll 2g/\Delta_{g=0}^2$.

Due to the time dependence of optomechanical coupling, the eigenmodes also attain time dependence. For clarity with the time evolution of modes, we denote them with time-dependent coupling as $A_{k,g}(t)$ and $B_{k,g}(t)$. To investigate the nonadiabatic dynamics of the populations, $\langle A_{k,g}^\dagger A_{k,g}(t)\rangle$ and $\langle B_{k,g}^\dagger B_{k,g}(t)\rangle$, of the eigenmodes, we first compute the initial populations, $\langle A_{k,g}^\dagger A_{k,g}(0)\rangle$ and $\langle B_{k,g}^\dagger B_{k,g}(0)\rangle$ [see appendix B] by assuming that prior to any change in the Hamiltonian, the system attains the equilibrium state with its environment. Then, the time evolution of $\langle A_{k,g}^\dagger A_{k,g}(t)\rangle$ and $\langle B_{k,g}^\dagger B_{k,g}(t)\rangle$ is obtained by computing the time propagator $S_k(g(t))$ from the equations of motion for $a_k(g(t))$ and $b_k(g(t))$ [see appendix A] using Magnus expansion method [33]. The propagator reads as:

$$S_{k,g}(t) = \begin{pmatrix} cos\eta(t) - i\frac{\phi(t)}{\eta(t)}sin\eta(t) & i\frac{\theta(t)}{\eta(t)}sin\eta(t) \\ i\frac{\theta^*(t)}{\eta(t)}sin\eta(t) & cos\eta(t) + i\frac{\phi(t)}{\eta(t)}sin\eta(t) \end{pmatrix}, \tag{5}$$

$$\begin{pmatrix} a_{k,g}(t) \\ b_{k,g}(t) \end{pmatrix} = S_{k,g}(t)\begin{pmatrix} a_k \\ b_k \end{pmatrix} \tag{6}$$

where, $\eta(t) = \sqrt{|\theta(t)|^2 + \phi^2(t)}$, $\theta(t) = \frac{g}{2\delta}\left[i\left(e^{2i\delta t} - 1\right) - \frac{2ite^{2i\delta t}}{t_q} + \frac{1}{\delta t_q}\left(e^{2i\delta t} - 1\right)\right]$, and $\phi(t) = \frac{g^2}{4t_q\delta^3}\xi + \frac{g^2}{t_q\delta}\zeta + \frac{g^2}{2\delta}\chi$. Here, $\xi = \left[1 - cos(2\delta t) + \frac{2t}{t_q}cos(2\delta t) - \frac{1}{t_q\delta}sin(2\delta t)\right]$, $\zeta = \left[\frac{t^2}{2} - \frac{2t^3}{3t_q}\right]$, and $\chi = \left[t - \frac{t^2}{t_q} - \frac{sin(2\delta t)}{2\delta} + \frac{1}{t_q\delta}tsin(2\delta t) + \frac{1}{2t_q\delta^2}\{cos(2\delta t) - 1\}\right]$. We use $S_{k,g}(t)$, $a_k$ and $b_k$ for calculating the temporal dynamics of eigen modes as follows:

$$\begin{pmatrix} A_{k,g}(t) \\ B_{k,g}(t) \end{pmatrix} = R_{k,g}(t)S_{k,g}(t)\begin{pmatrix} a_k \\ b_k \end{pmatrix} \tag{7}$$

$$\begin{pmatrix} A_{k,g}(t) \\ B_{k,g}(t) \end{pmatrix} = R_{k,g}(t)S_{k,g}(t)R_k^{-1}\begin{pmatrix} A_k \\ B_k \end{pmatrix}, \tag{8}$$

The populations of eigenmodes, $N(A_k) = \langle A_{k,g}^\dagger A_{k,g}(t)\rangle / n_{th}$ and $N(B_k) = \langle B_{k,g}^\dagger B_{k,g}(t)\rangle / n_{th}$, are calculated using Eqs. (5) and (8), here $n_{th}$ is the average number of thermal phonons. The net excitations in the eigenmodes are defined as $N_q(A_k) = \left(\langle A_{k,g}^\dagger A_{k,g}(t)\rangle - \langle A_{k,g}^\dagger A_{k,g}(0)\rangle\right)/n_{th}$ and $N_q(B_k) = \left(\langle B_{k,g}^\dagger B_{k,g}(t)\rangle - \langle B_{k,g}^\dagger B_{k,g}(0)\rangle\right)/n_{th}$.

We show the nonadiabatic dynamics of eigenmodes in Figs. 2 and 3. It can be observed from Fig. 2(a) that for $kd = 0.48\pi$ and $\theta = 0$, initially, the population of eigenmode $A_{k,g}$ is much greater, however in contrast to the adiabatic dynamics, the transfer of population occurs from mode $A_{k,g}$ to mode $B_{k,g}$ over the quench time, indicating the excitation of mode $B_{k,g}$. Remarkably, for the identical value of $kd$ but for a different phase, $\theta = \pi$, the scenario is entirely inverted; specifically, the initial population of eigenmode $B_{k,g}$ is significantly larger, resulting in excitation transferring from mode $B_{k,g}$ to mode $A_{k,g}$, as depicted in Fig. 2(b). Next in Fig. 2(c), the net excitations in eigenmodes are shown for $\theta = \pi$. The nonadiabatic dynamics is evident from the generation of net excitations in the eigenmodes around $kd = \pm\pi/2$. Notably at $kd = \pi/2(-\pi/2)$, the transfer of net excitations takes place from mode $B_{k,g}\left(A_{k,g}\right)$ to mode $A_{k,g}\left(B_{k,g}\right)$. The similar trend of excitations can be observed in Fig 2(d) for $\theta = 0.8\pi$; however, the dominant excitations are now shifted to $kd = -0.34\pi$ and $0.66\pi$. This clearly demonstrates how the phase of the laser drive influences population dynamics and net excitations in the eigenmodes. Prior to elucidating the underlying physics in the subsequent part, we demonstrate the net excitations in the eigenmodes in a contrasting parameter regime, wherein the optomechanical coupling exceed the hopping strengths, $J, K < g$. It is clear from Figs. 3(a) and 3(d) that for $\theta = 0$ and $\pi$, the quick transfer of net excitations occurs at $kd = \pm\pi/2$.

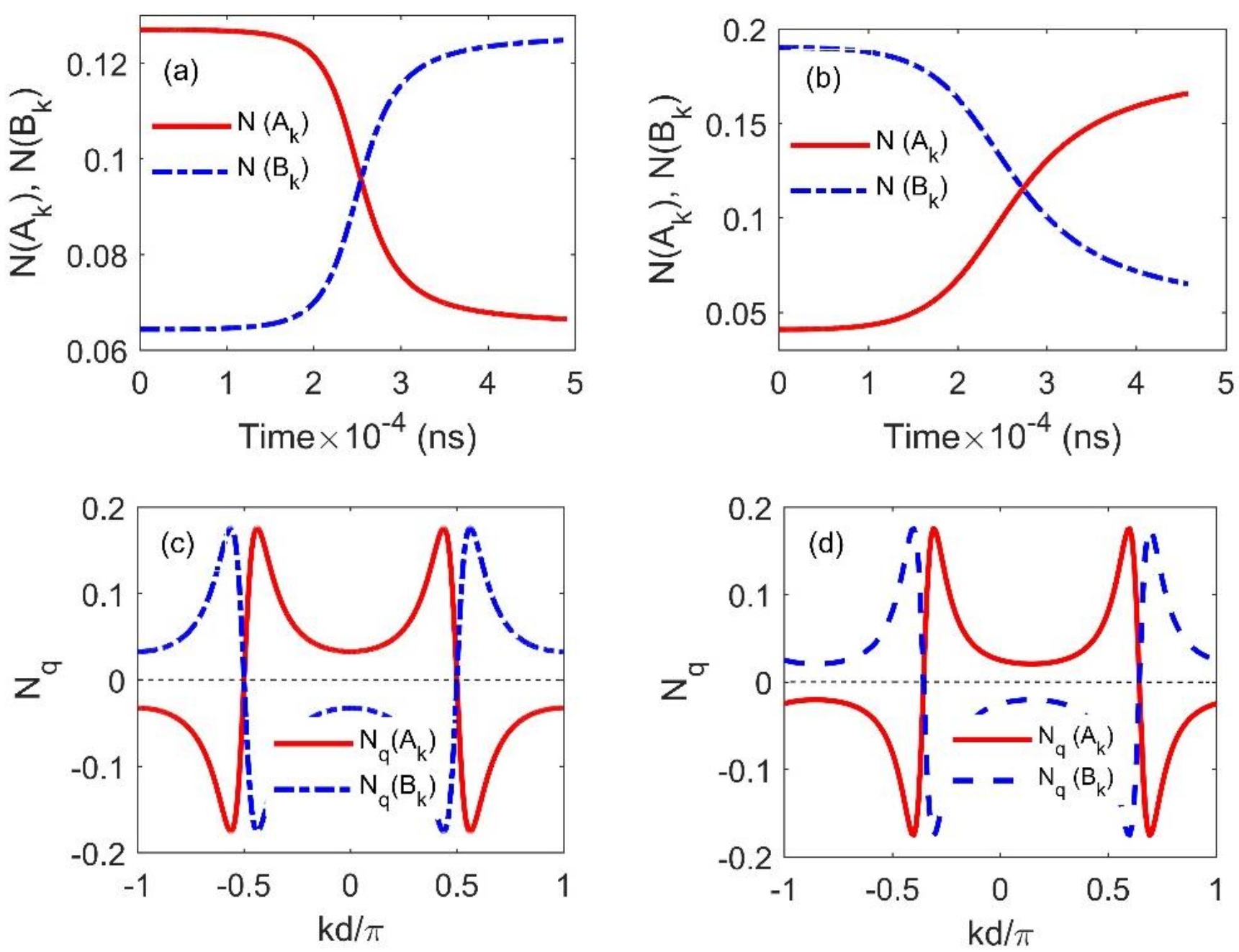

Fig. 2 (color online) Dynamics of eigenmodes due to the switching of optomechanical coupling from $+g\ to - g$ over a quench time $t_q = 10^{-4}g/\Delta_{g=0}^2$. (a) $kd = 0.48\pi$ and $\theta = 0$, (b) $kd = 0.48\pi$ and $\theta = \pi$. Net excitation in eigen modes: (c) $\theta = \pi$, and (d) $\theta = 0.8\pi$. The values of other simulation parameters are given as $\omega_m = 4.3\ GHz$, $J = 0.5\ GHz$, $g = 0.1\ GHz$, and $K = 0.2\ GHz$.

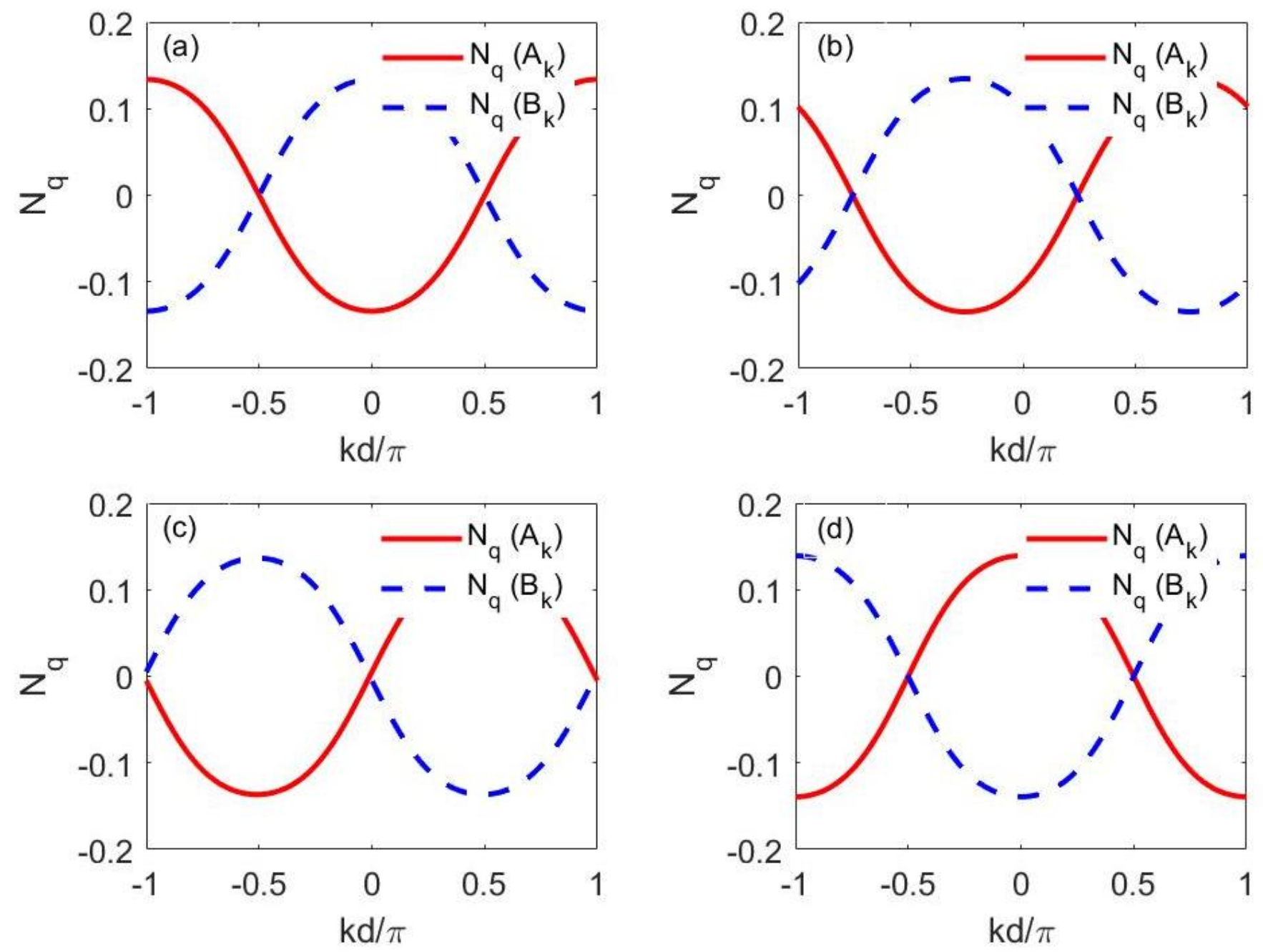

Fig. 3 (color online) Net excitations in the eigen modes for $t_q = 10^{-4}g/\Delta_{g=0}^2$ (a) $\theta = 0$, (b) $\theta = \pi/4$, (c) $\theta = \pi/2$, and (d) $\theta = \pi$. The values of other simulation parameters are given as $g = 0.086\ GHz, \omega_m = 4.3\ GHz$, $J = 0.043\ GHz$, and $K = 0.0013\ GHz$.

However, for $\theta = 0$, the transfer of net excitations takes place from mode $A_{k,g}(B_{k,g})$ to mode $B_{k,g}(A_{k,g})$ at $kd = -\pi/2\ (\pi/2)$, while exactly opposite transfer of net excitations occurs for $\theta = \pi$. For $\theta = \pi/4$, the fast transfer of net excitations gets shifted to $kd = -0.75\pi$ and $0.25\pi$. Similarly, Fig. 3(c) illustrates the phase-induced shifting of fast transfer of net excitations between the eigen modes to $kd = 0$. To clarify the role of the phase-dependent alteration of the rapid transfer of net excitations among the eigenmodes and the transition of net excitations between the eigenmodes due to a phase change, we will further examine the band structure of the eigenmodes.

**B. Control of band structure and eigenmodes**

We investigate the phase-induced modulation of the band structure of eigenmodes to clarify its significance in the nonadiabatic dynamics of these modes. Figure 4 illustrates the dependence of the energy gap on the normalized wavenumber for various values of $g$ and $\theta$. It is clear from Fig. 4 (a) that the energy gap between the two modes is maximum at $kd = 0$, while it vanishes at the two crossings $kd = \pm\pi/2$. However, for $g = 0.1\ GHz$ in Fig. 4(b), the energy gap opens up with avoided crossings at $kd = \pm\pi/2$.

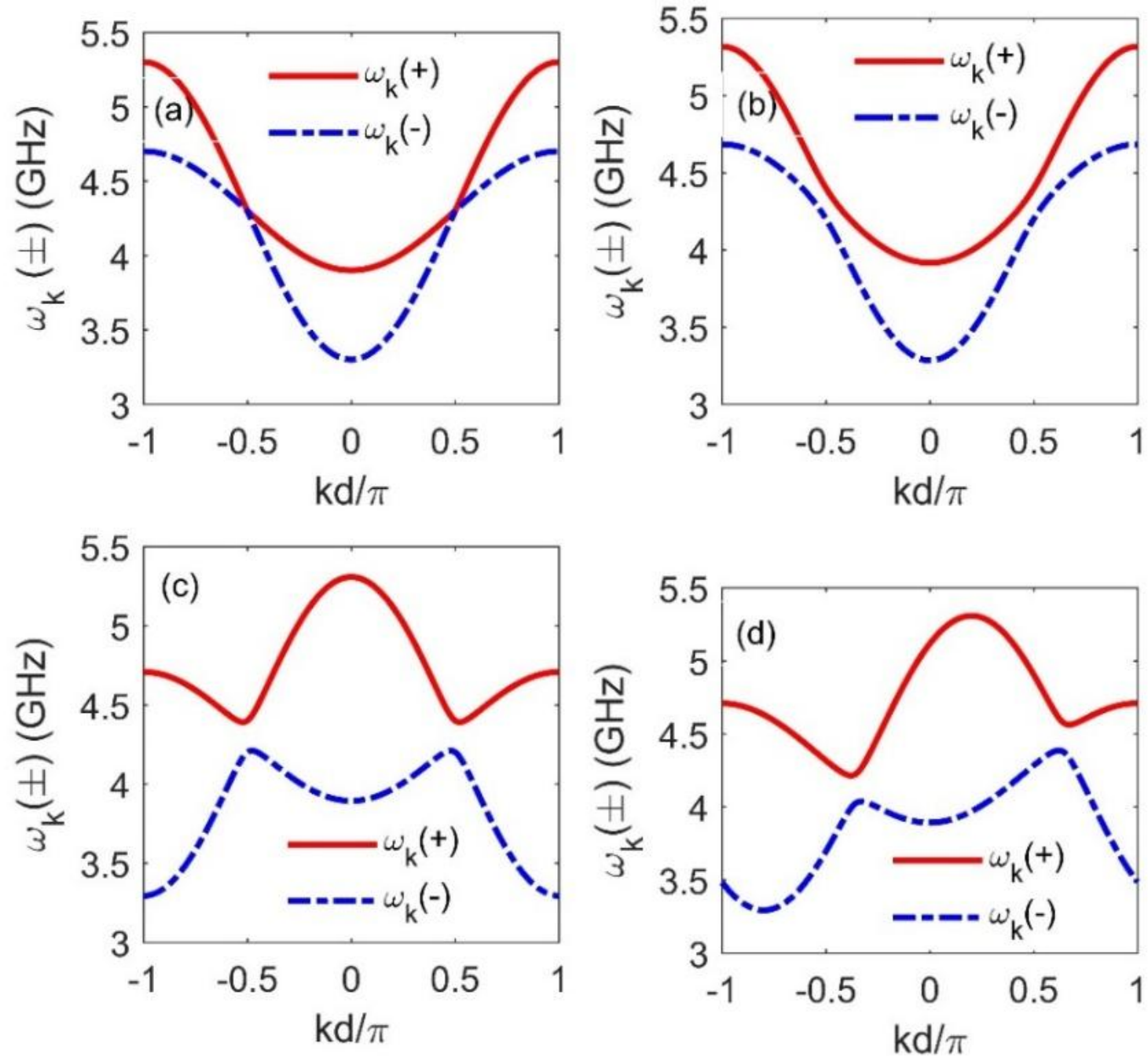

Fig. 4 (color online) Energy of the hybrid eigenmodes as a function of normalized wavenumber for detuning $\Delta = -\omega_m$. (a) $g = 0$ and $\theta = 0$, (b) $g = 0.1\ GHz$ and $\theta = 0$, (c) $g = 0.1\ GHz$ and $\theta = \pi$, and (d) $g = 0.1\ GHz$ and $\theta = 0.8\pi$. The values of other simulation parameters are identical to those used in Fig. 2.

Interestingly, for a nonzero value of the phase of the laser drive, the band structure changes significantly, as illustrated in Figs. 4(c) and 4 (d). For $\theta = \pi$, the energy gap is minimum at $kd = \pm\pi/2$. This is why the quick transfer of net excitations takes place in the vicinity of these values of $kd$ in Fig. 2(c).

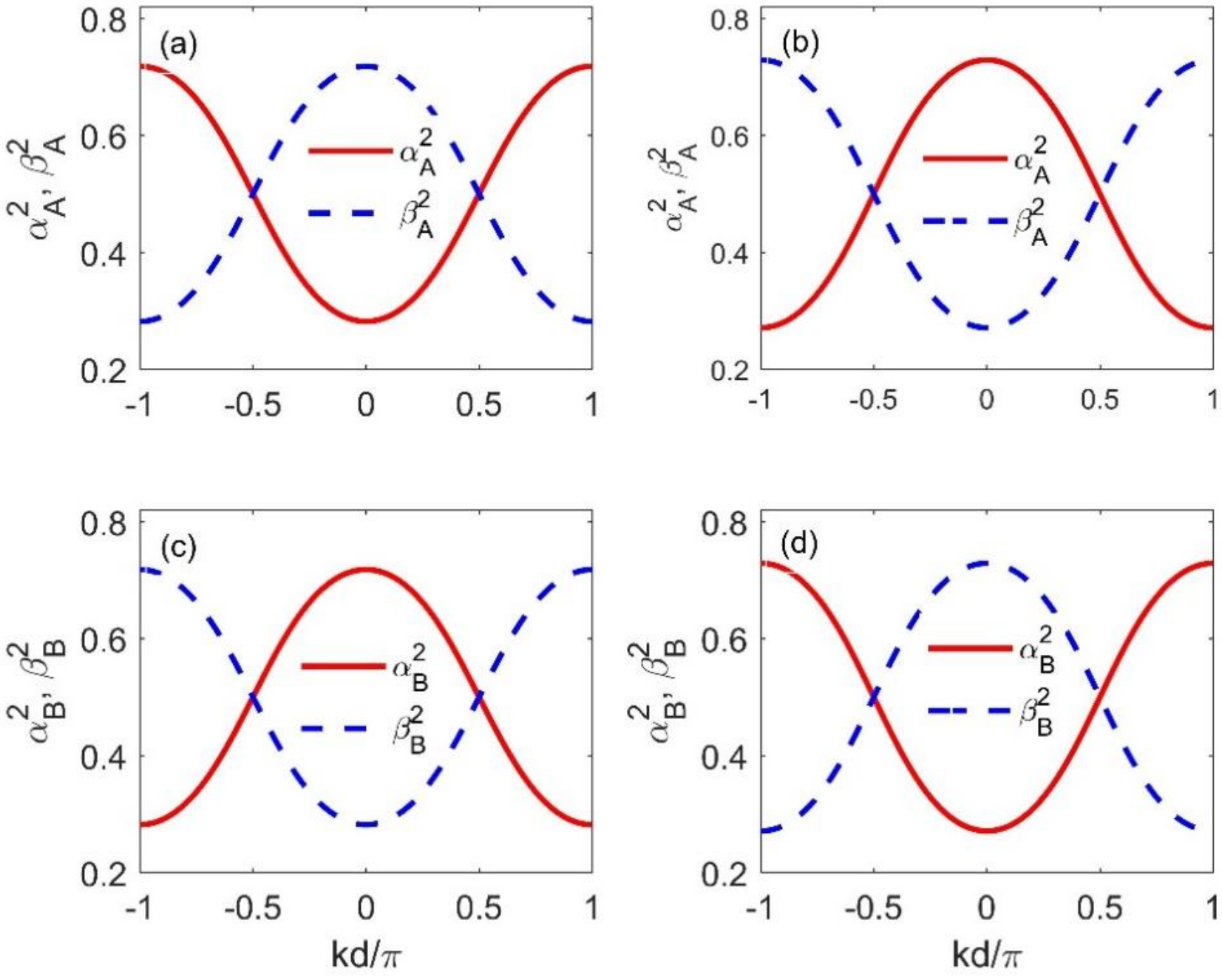

Fig. 5 (color online) The relative weights of photons and phonons composing the hybrid eigenmodes. (a) $\alpha_A^2$ and $\beta_A^2$ for $\theta = 0$, (b) $\alpha_A^2$and $\beta_A^2$ for $\theta = \pi$, (c) $\alpha_B^2$ and $\beta_B^2$ for and $\theta = 0$, and (d) $\alpha_B^2$ and $\beta_B^2$ for $\theta = \pi$. The values of other simulation parameters are identical to those employed in Fig. 3.

However, for $\theta = 0.8\pi$, minimum energy gaps become asymmetric and shifts to $kd = 0.66\pi$ and $-0.34\pi$. This phase-induced shifting of the minimum energy gaps is a primary reason for shifting the sharp transfer of net excitations, as shown in Fig. 2(d). Furthermore, to clarify the switching of the net excitations between the eigenmodes for a phase change from $\theta = 0$ to $\pi$ as illustrated in Figs. 3(a) and 3(d), we show the relative weight of photons and phonons composing the hybrid eigenmodes, $A_k$ and $B_k$ at two distinct phases, $\theta = 0$ and $\pi$. The eigenmode $A_k$ is mostly phononic in the middle, $-\pi/2 < kd < \pi/2$, and outside it is mostly photonic, as can be observed in Fig. 5(a). However, the switching of the relative weights of photons and phonons do occur for $\theta = \pi$, specifically, $A_k$ becomes mostly photonic in the middle, $-\pi/2 < kd < \pi/2$, and outside it is mostly phononic. Similar switching of the relative weights happens in eigenmode $B_k$ for $\theta = \pi$, as illustrated in Figs. 5(c) and 5(d). Therefore, the switching of the net excitations between the eigenmodes can be attributed to the transformation of the eigenmode $A_k(B_k)$ into $B_k(A_k)$ as the phase changes from $\theta = 0$ to $\pi$.

To clarify the phase-dependent shifting of the transfer of net excitations illustrated in Fig. 3, we now use the identical parameters and investigate the band structure of the eigenmodes. As expected, it can be observed from Fig.6 that the shifting of energy gap minima depends on the value of $\theta$. Specifically, the lowest energy gaps appear at $kd = \pm\pi/2$ for $\theta = 0$ and $\pi$, while they shift to $kd = -0.75\pi$ and $0.25\pi$ for $\theta = \pi/4$. Similar shifting to $kd = 0$ for $\theta = \pi/2$ can also be observed in Fig. 6 (c). Clearly, the phase-induced shifting of lowest energy gaps enables the shifting of fast transfer of net excitations as illustrated in Fig. 3.

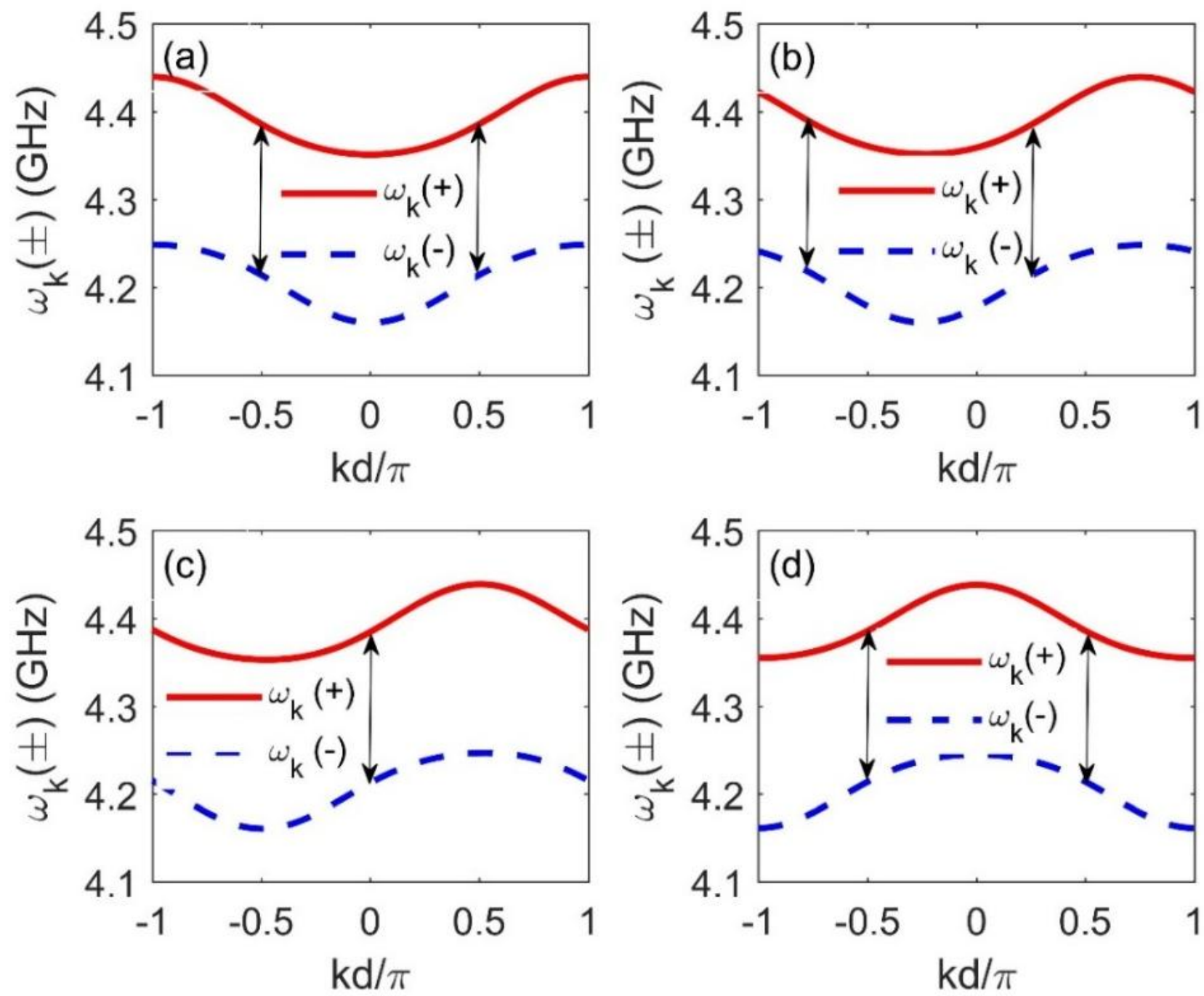

Fig. 6 (color online) Energy of the hybrid eigenmodes as a function of normalized wavenumber for detuning $\Delta = -\omega_m$. (a) $\theta = 0$, (b) $\theta = \pi/4$, (c) $\theta = \pi/2$, and (d) $\theta = \pi$. The double side arrows indicate the minimum gap. The values of other simulation parameters are identical to those used in Fig. 3.

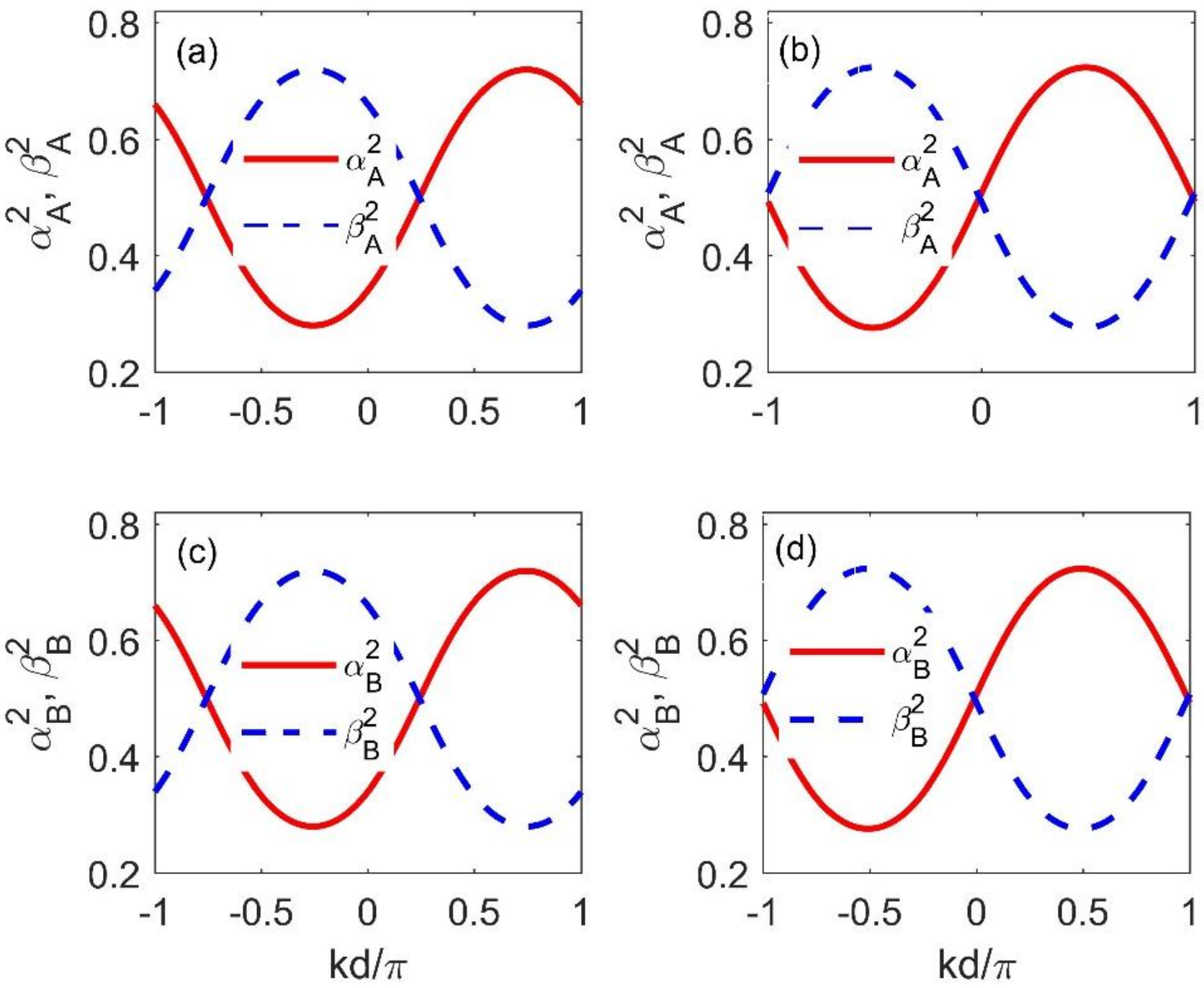

Fig. 7 (color online) The relative weights of photons and phonons in hybrid eigenmodes. (a) $\alpha_A^2$ and $\beta_A^2$ for $\theta = \pi/4$, (b) $\alpha_A^2$and $\beta_A^2$ for $\theta = \pi/2$, (c) $\alpha_B^2$ and $\beta_B^2$ for $\theta = \pi/4$, and (d) $\alpha_B^2$ and $\beta_B^2$ for $\theta = \pi/2$. The value of other simulation parameters is same as employed in Fig. 3.

Moreover in Fig. 7, we illustrate the relative weights of photons and phonons to clarify the role of hybridization for two different phases. It can be understood from Figs. 6(b), 7(a), and 7(c) that the energy gap between the hybrid modes is smallest when the photonic and phononic modes contribute equally, leading to strong excitations in the eigenmodes [see Fig. 3(b)]. Similar inferences can be drawn from Figs. 6(c), 7(b), and 7(d). Additionally equal contributions of photon and phonon modes appear at $kd = -0.75\pi$ and $0.25\pi$ for $\theta = \pi/4$ and $kd = 0$ for $\theta = \pi/2$, respectively, leading to the generation of net excitations at these values of $kd$ [see Figs. 3(b) and 3(c)]. Here, we note that the observed phase-dependent degree of hybridization and net excitations in the eigenmodes may be probed experimentally by analysing the transmission of a photon or phonon field in the optomechanical array with input and output coupling contacts [17, 30].

### C. Coherent photon-phonon conversion

Finally, we investigate the phase-dependent photon-phonon conversion in a four- and five-site arrays by numerically solving the quantum master equation in Lindblad form, which is given in appendix B. In Figs. 8 (a) and 8(b), we show the dynamics of photon conversion at the first and forth sites by assuming that initially forth and first phonon sites are occupied, i.e., $|\psi\rangle_{b_4} = |0\rangle_{a_1} \otimes |0\rangle_{b_1} \otimes |0\rangle_{a_2} \otimes |0\rangle_{b_2} \dots \dots |0\rangle_{a_4} \otimes |1\rangle_{b_4}$ and $|\psi\rangle_{b_1} = |0\rangle_{a_1} \otimes |1\rangle_{b_1} \otimes |0\rangle_{a_2} \otimes |0\rangle_{b_2} \dots \dots |0\rangle_{a_4} \otimes |0\rangle_{b_4}$, respectively [8]. Similarly, Figs. 8(c) and 8(d) illustrate the phonon conversion for initially occupied photonic modes at forth and first sites, i.e., $|\psi\rangle_{a_4} = |0\rangle_{a_1} \otimes |0\rangle_{b_1} \otimes |0\rangle_{a_2} \otimes |0\rangle_{b_2} \dots \dots |1\rangle_{a_4} \otimes |0\rangle_{b_4}$ and $|\psi\rangle_{a_1} = |1\rangle_{a_1} \otimes |0\rangle_{b_1} \otimes |0\rangle_{a_2} \otimes |0\rangle_{b_2} \dots \dots |0\rangle_{a_4} \otimes |0\rangle_{b_4}$, respectively. In principle, photonic (phononic) sites can be prepared in a single photon (phonon) state through optomechanical and piezoelectric couplings of the optomechanical array to the hybrid quantum systems [8, 34-36].

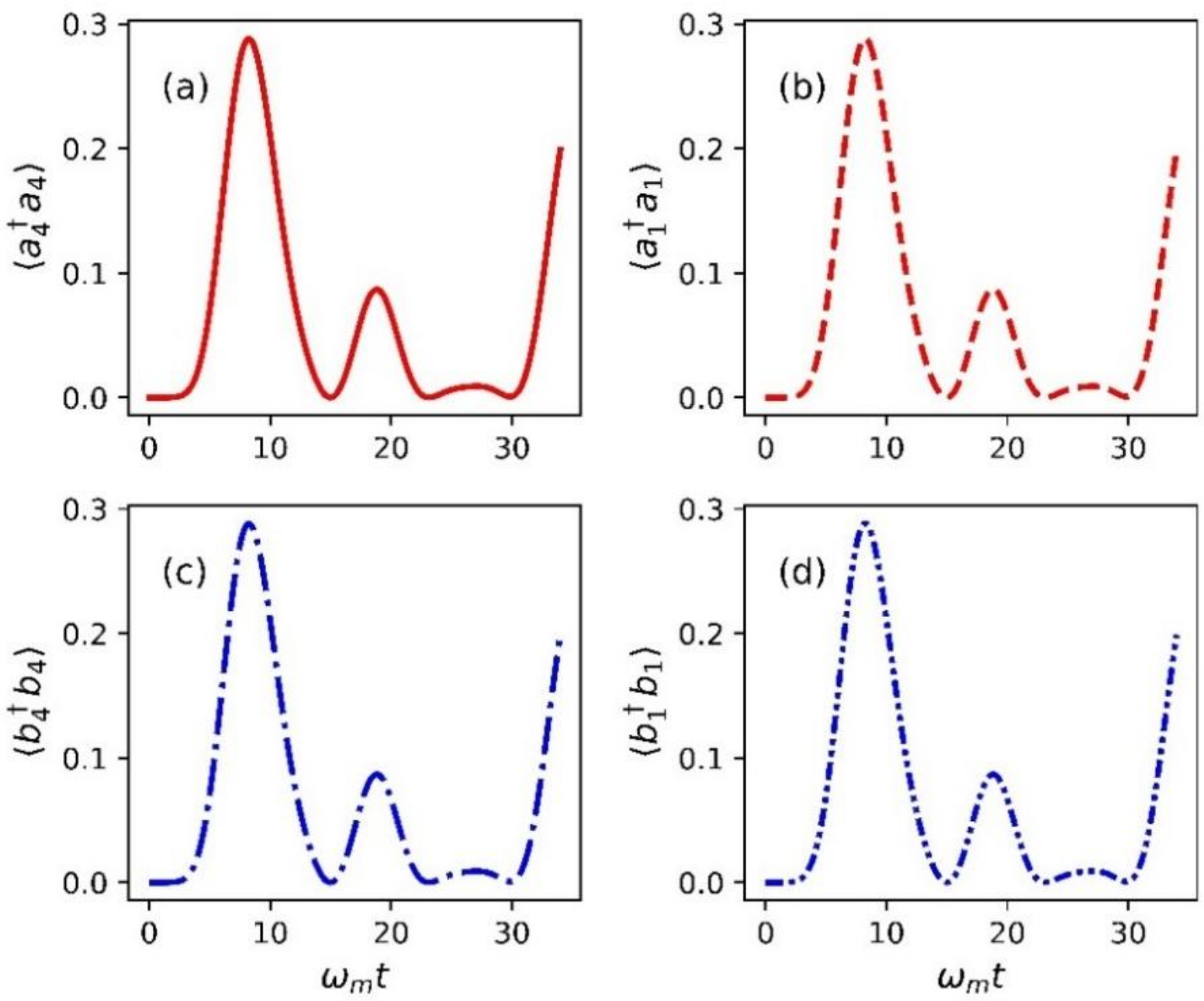

Fig. 8 (color online) dynamics of photon and phonon populations for the initially prepared phononic and photonics states: (a) $|\psi\rangle_{b_1}$, (b) $|\psi\rangle_{b_4}$, (c) $|\psi\rangle_{a_1}$, and (d) $|\psi\rangle_{a_4}$ at $\kappa = 0.01\omega_m$, $\Gamma = 0.001\omega_m$, and $\theta = 0$. The other parameters are same as those used in Fig. 2.

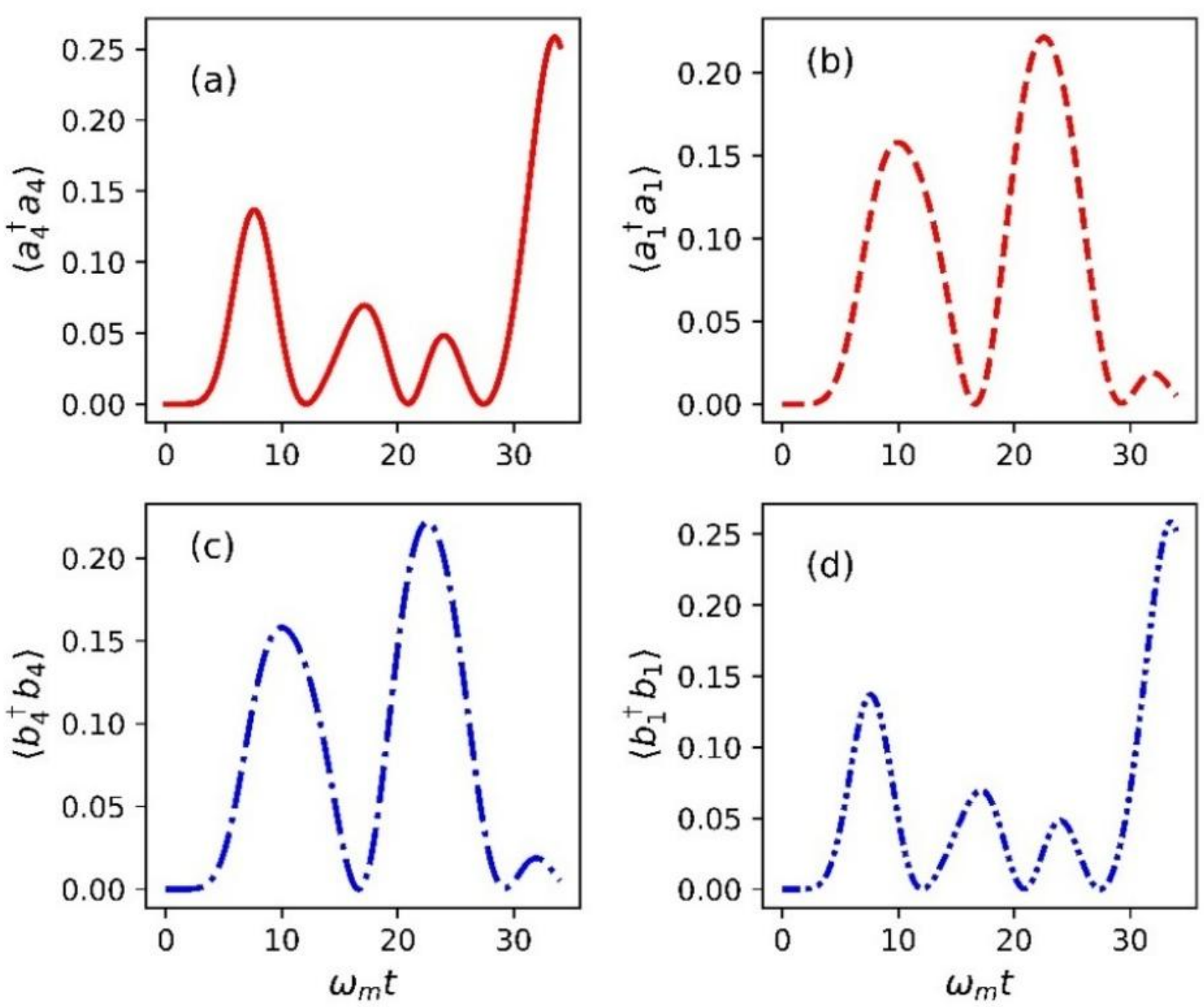

Fig. 9 (color online) dynamics of photon and phonon populations for the initially prepared phononic and photonics states: (a) $|\psi\rangle_{b_1}$, (b) $|\psi\rangle_{b_4}$, (c) $|\psi\rangle_{a_1}$, and (d) $|\psi\rangle_{a_4}$ at $\kappa = 0.01\omega_m$, $\Gamma = 0.001\omega_m$, and $\theta = \pi$. The other parameters are same as those used in Fig. 2.

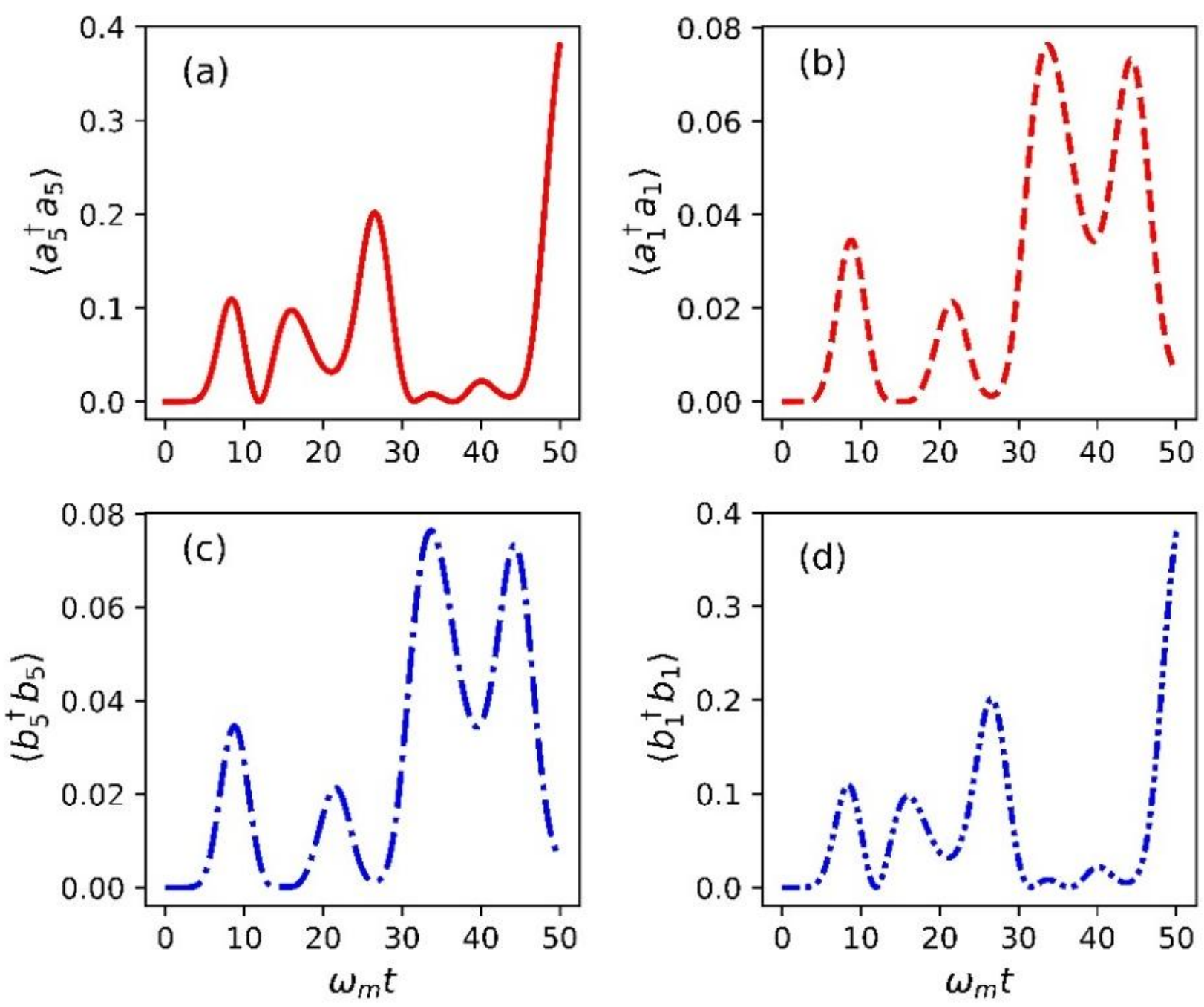

Fig. 10 (color online) dynamics of photon and phonon populations in a five-site array for the initially prepared phononic and photonics states: (a) $|\psi\rangle_{b_1}$, (b) $|\psi\rangle_{b_5}$, (c) $|\psi\rangle_{a_1}$, and (d) $|\psi\rangle_{a_5}$ at $\kappa = 0.01\omega_m$, $\Gamma = 0.001\omega_m$, and $\theta = \pi$. The other parameters are same as those used in Fig. 2.

It is clear from Fig. 8 that for phase $\theta = 0$, the dynamics of photon and phonon populations are identical irrespective of the initial photonic and phononic states. However, it is evident from Fig. 9 that for phase $\theta = \pi$, the dynamics of photon and phonon populations are no longer identical. Specifically, for an initially populated first phononic site $|\psi\rangle_{b_1}$, the average photon number at fourth site is maximum at $\omega_m t = 33$, while it is minimum at first site if the initial state is $|\psi\rangle_{b_1}$. Figures 9 (c) and (d) exhibit analogous trends in the phase-dependent conversion of photons to phonons. Consequently, the conversion and propagation of photons (phonons) between the first and fourth sites exhibit a pronounced phase dependence. The maximum efficiency of photon-phonon conversion for the selected parameters is approximately 26%, as illustrated in Figs. 9(a) and 9(d).

We now show the photon-phonon conversion in a five-site array for $\theta = \pi$. Like Fig. 9, the conversion and propagation of photons (phonons) between the first and fifth sites shows a strong dependence on the phase of the driving laser. For an initially populated first phononic site, $|\psi\rangle_{b_1} = |0\rangle_{a_1} \otimes |1\rangle_{b_1} \otimes |0\rangle_{a_2} \otimes |0\rangle_{b_2} \dots \dots |0\rangle_{a_5} \otimes |0\rangle_{b_5}$, the average photon number at fifth site is maximum at $\omega_m t = 50$, while it is minimum at first site for $|\psi\rangle_{b_5} = |0\rangle_{a_1} \otimes |0\rangle_{b_1} \otimes |0\rangle_{a_2} \otimes |0\rangle_{b_2} \dots \dots |0\rangle_{a_5} \otimes |1\rangle_{b_5}$. The conversion of photons into phonons follows a similar trend as shown in Figs. 10 (c) and (d). In contrast to the four-site array, the maximum efficiency of photon-phonon conversion in a five-site array is improved, exceeding 38%, as demonstrated in Fig. 10(a) and 10(d). We expect that the phase dependence of photon-phonon conversion results from the interference among the several pathways that a converted photon (phonon) may take to arrive at the first or fifth site [30]. Nonetheless, further analytical and simulation investigations are necessary to validate this.

## IV. Conclusion

In conclusion, we have investigated the impact of the phase of a laser drive on the nonadiabatic dynamics of eigenmodes and the coherent conversion of photon-phonons in an array of optomechanical cavities. We have shown that the phase gradient of the driving laser alters and shifts the net excitations in eigenmodes. The band structure of eigenmodes elucidates that the shift of net excitations arises from phase-induced alterations in the energy band gaps, whereas the transition of net excitations in eigenmodes occurs as the phase varies from zero to $\pi$ due to the switching of photonic and phononic weights in eigenmodes. Additionally, we have examined the influence of phase on photon-phonon conversion in a four- and five-site array, with the latter configuration exhibiting superior conversion efficiency. These findings may foster further studies to investigate the role of phase-dependent nonadiabatic dynamics in quantum phase transitions and topologically protected edge states.

**Acknowledgements**

P. K. gratefully acknowledges the new faculty seed grant from the Indian Institute of Technology Delhi.

## APPENDIX A: EQUATION OF MOTIONS FOR PHOTONIC AND PHONONIC MODES

The time dependent Hamiltonian in Fourier basis is given as:

$$H_k(t) = \begin{pmatrix} -\Delta(k) & -g(t) \\ -g(t) & \Omega(k) \end{pmatrix} \quad \text{(A1)}$$

We consider the following ansatz wavefunction:

$$\psi_k(t) = a_k\big(g(t)\big)e^{i\Delta(k)t}\psi_a + b_k\big(g(t)\big)e^{-i\Omega(k)t}\psi_b, \quad \text{(A2)}$$

here, $\Delta(k) = \Delta + 2J\,cos(kd + \theta)$, $\Omega(k) = \omega_m - 2Kcos(kd)$, $a_k(t)$, and $b_k(t)$ are the probability amplitudes of the photonic and phononic modes. To derive the equation of motion we use the time dependent Schrödinger equation.

$$i\hbar\frac{d\psi_k(t)}{dt} = H_k(t)\psi_k(t) \quad \text{(A3)}$$

From Eqs. A1, A2, and A3, we get the following equations of motion:

$$\dot{a}_k\big(g(t)\big) = ig(t)b_k\big(g(t)\big)e^{-i\Delta'(k)t} \quad \text{(A4)}$$

$$\dot{b}_k\big(g(t)\big) = ig(t)a_k\big(g(t)\big)e^{i\Delta'(k)t}, \quad \text{(A5)}$$

where, $\Delta'(k) = \Delta(k) + \Omega(k)$.

The eigen modes population, $\langle A^\dagger_{k,g}A_{k,g}(t)\rangle$ and $\langle B^\dagger_{k,g}B_{k,g}(t)\rangle$, is calculated with the aid of Eqs. 5 and 8 in the main text and given as follows:

$$\langle A^\dagger_{k,g}A_{k,g}(t)\rangle = \frac{1}{(\alpha_A\beta_B - \alpha_B\beta_A)^2}\big[|A_1|^2\langle A^\dagger_{k,g}A_{k,g}(0)\rangle + |B_1|^2\langle B^\dagger_{k,g}B_{k,g}(0)\rangle\big] \quad \text{(A6)}$$

$$\langle B^\dagger_{k,g}B_{k,g}(t)\rangle = \frac{1}{(\alpha_A\beta_B - \alpha_B\beta_A)^2}\big[|A_2|^2\langle A^\dagger_{k,g}A_{k,g}(0)\rangle + |B_2|^2\langle B^\dagger_{k,g}B_{k,g}(0)\rangle\big], \quad \text{(A7)}$$

$$A_1 = \beta_B[\alpha_A\{g(t)\}s_{11} + \beta_A\{g(t)\}s_{21}] - \alpha_B[\alpha_A\{g(t)\}s_{12} + \beta_A\{g(t)\}s_{22}], \quad \text{(A8)}$$

$$B_1 = \alpha_A[\alpha_A\{g(t)\}s_{12} + \beta_A\{g(t)\}s_{22}] - \beta_A[\alpha_A\{g(t)\}s_{11} + \beta_A\{g(t)\}s_{21}], \quad \text{(A9)}$$

$$A_2 = \beta_B[\alpha_B\{g(t)\}s_{11} + \beta_B\{g(t)\}s_{21}] - \alpha_B[\alpha_B\{g(t)\}s_{12} + \beta_B\{g(t)\}s_{22}], \quad \text{(A10)}$$

$$B_2 = \alpha_A[\alpha_B\{g(t)\}s_{12} + \beta_B\{g(t)\}s_{22}] - \beta_A[\alpha_B\{g(t)\}s_{11} + \beta_B\{g(t)\}s_{21}], \quad \text{(A11)}$$

Here, $s_{11}$, $s_{12}$, $s_{21}$, and $s_{22}$ are the matrix elements of $S_{k,g}(t)$, given by Eq. 5 in the main text, while $\langle A^\dagger_{k,g}A_{k,g}(0)\rangle$ and $\langle B^\dagger_{k,g}B_{k,g}(0)\rangle$ are the initial populations of the eigenmodes.

**APPENDIX B: STEADY STATE POPULATION OF EIGEN MODES**

The initial populations of the eigenmodes, $\langle A^\dagger_{k,g}A_{k,g}(0)\rangle$ and $\langle B^\dagger_{k,g}B_{k,g}(0)\rangle$, correspond to the steady state populations of the eigenmodes of the Hamiltonian with a time-independent optomechanical coupling $g$. We follow the Ref. [18] for deriving the steady state population of eigenmodes. The equations of motion for $a_k(t)$ and $b_k(t)$ with constant optomechanical coupling strength is derived using Hamiltonian given by Eq. 4 and employing Heisenberg-Langevin equations, which reads as:

$$\frac{d}{dt}\begin{pmatrix} a_k \\ b_k \end{pmatrix} = \begin{pmatrix} i\Delta(k) - \frac{\kappa}{2} & ig \\ ig & -i\Omega(k) - \frac{\Gamma}{2} \end{pmatrix}\begin{pmatrix} a_k \\ b_k \end{pmatrix} + \begin{pmatrix} \sqrt{\kappa} a_{in} \\ \sqrt{\Gamma} b_{in} \end{pmatrix}, \tag{B1}$$

here, $\kappa$ and $\Gamma$ are the decay rates of photonic and phononic modes, while $a_{in}$ and $b_{in}$ are the input noise operators of photonic and phononic modes. For $g = 0$, it is straightforward to find the solution of $b_k(t)$ from Eq. A6. It reads as

$$b_k(t) = e^{-\left(i\Omega(k) + \frac{\Gamma}{2}\right)t}\left[b_k(0) + \sqrt{\Gamma}\int_0^t dt'\, e^{-\left(i\Omega(k) + \frac{\Gamma}{2}\right)t'} b_{in}(t')\right] \tag{B2}$$

The Heisenberg-Langevin equations of motion for eigen modes, $A_k$ and $B_k$ reads as follows:

$$\frac{d}{dt}\begin{pmatrix} A_k \\ B_k \end{pmatrix} = \begin{pmatrix} -i\omega_k(+) - \frac{\kappa_A}{2} & 0 \\ 0 & -i\omega_k(-) - \frac{\kappa_B}{2} \end{pmatrix}\begin{pmatrix} A_k \\ B_k \end{pmatrix} + \begin{pmatrix} A_{in} \\ B_{in} \end{pmatrix} \tag{B3}$$

$\kappa_A$ and $\kappa_B$ represent the decay rates, while $A_{in}$ and $B_{in}$ are the linear superposition of $a_{in}$ and $b_{in}$. We calculate the $\kappa_A$ and $\kappa_B$ as the first-order perturbation to matrix $M$, given below, without decay rates.

$$M = \begin{pmatrix} i\Delta(k) & ig \\ ig & -i\Omega(k) \end{pmatrix} + \begin{pmatrix} -\frac{\kappa}{2} & 0 \\ 0 & -\frac{\Gamma}{2} \end{pmatrix} \tag{B4}$$

From Eq. A8, the population of eigen modes $A_k$ and $B_k$ can be calculated similar to the isolated mode $b_k$ in Eq. A7. Following the calculations, we get

$$\langle A_k^\dagger A_k(t)\rangle = e^{-\kappa_A t}\left[\langle A_k^\dagger A_k(0)\rangle + \int_0^t\int_0^t dt'\, dt'' e^{\kappa_A(t'+t'')/2} e^{\omega_k(+)(t'-t'')}\langle A_{in}^\dagger(t') A_{in}(t'')\rangle\right] \tag{B5}$$

$$\begin{pmatrix} A_{in}(t) \\ B_{in}(t) \end{pmatrix} = \begin{pmatrix} \alpha_A & \beta_A \\ \alpha_B & \beta_B \end{pmatrix}\begin{pmatrix} \sqrt{\kappa} a_{in}(t) \\ \sqrt{\Gamma} b_{in}(t) \end{pmatrix} \tag{B6}$$

From Eq. A11, $\langle A_{in}^\dagger(t') A_{in}(t'')\rangle$ is calculated and reads as follows:

$$\langle A_{in}^\dagger(t') A_{in}(t'')\rangle = \alpha_A^2 \kappa \langle a_{in}^\dagger(t') a_{in}(t'')\rangle + \beta_A^2 \Gamma \langle b_{in}^\dagger(t') b_{in}(t'')\rangle, \tag{B7}$$

here, $\alpha_A^2 + \beta_A^2 = 1$. Under Markov approximation with optical bath at zero temperature, we get

$$\langle a_{in}^\dagger(t') a_{in}(t'')\rangle = 0, \langle b_{in}^\dagger(t') b_{in}(t'')\rangle = n_{th}\delta(t' - t''), \tag{B8}$$

here, $n_{th}$ is the average number of thermal phonons.

From Eqs. A10, A12, and A13, we get,

$$\langle A_k^\dagger A_k(t)\rangle = e^{-\kappa_A t}\Big[\langle A_k^\dagger A_k(0)\rangle + \int_0^t\int_0^t dt'\, dt'' e^{\kappa_A(t'+t'')/2} e^{\omega_k(+)(t'-t'')}\, \beta_A^2\Gamma n_{th}\delta(t'-t'')\Big] \quad \text{(B9)}$$

$$\langle A_k^\dagger A_k(t)\rangle = e^{-\kappa_A t}\Big[\langle A_k^\dagger A_k(0)\rangle + \int_0^t dt'\, e^{\kappa_A t'}\beta_A^2\Gamma n_{th}\Big] \quad \text{(B10)}$$

$$\langle A_k^\dagger A_k(t)\rangle = e^{-\kappa_A t}\left[\langle A_k^\dagger A_k(0)\rangle + \beta_A^2\Gamma n_{th}\left(\frac{e^{\kappa_A t}-1}{\kappa_A}\right)\right] \quad \text{(B11)}$$

$$\langle A_k^\dagger A_k(t)\rangle_{t\to\infty} = \frac{\beta_A^2\Gamma n_{th}}{\kappa_A}, \text{ where } \kappa_A = \alpha_A^2\kappa + \beta_A^2\Gamma$$

$$\langle A_k^\dagger A_k(t)\rangle_{t\to\infty} = \langle A_{k,g}^\dagger A_{k,g}(0)\rangle = \frac{(1-\alpha_A^2)\Gamma n_{th}}{\alpha_A^2\kappa + (1-\alpha_A^2)\Gamma} \quad \text{(B12)}$$

After repeating the same procedure, we get the steady state population of eigen mode $B_k$ as:

$$\langle B_k^\dagger B_k(t)\rangle_{t\to\infty} = \langle B_{k,g}^\dagger B_{k,g}(0)\rangle = \frac{\alpha_A^2\Gamma n_{th}}{(1-\alpha_A^2)\kappa + \alpha_A^2\Gamma} \quad \text{(B13)}$$

We use the following quantum master equation in Lindblad form for calculating the average population: $\langle a_n^\dagger a_n\rangle = tr(a_n^\dagger a_n\rho),:\langle b_n^\dagger b_n\rangle = tr(b_n^\dagger b_n\rho)$

$$\frac{d\rho}{dt} = -i[H_{eff},\rho] + \frac{\kappa}{2}\sum_n\left(2a_n\rho a_n^\dagger - a_n^\dagger a_n\rho - \rho a_n^\dagger a_n\right) + \frac{\Gamma}{2}\sum_n\left(2b_n\rho b_n^\dagger - b_n^\dagger b_n\rho - \rho b_n^\dagger b_n\right) \quad \text{(B14)}$$